# Black Holes LIGO/Virgo Domination and Single-lined Binaries with a Black Hole Candidate Component

Vladimir Lipunov[1,2], Evgeny Gorbovskoy[2], Valeria Grinshpun[1,2], Daniil Vlasenko[1,2]

[1] Lomonosov Moscow State University, Physics Department, 119991, Vorobievy hills, 1, Moscow, Russia
[2] Lomonosov Moscow State University, Sternberg Astronomical Institute, 119234, Universitetsky pr., 13, Moscow, Russia

ABSTRACT

In this letter, we note that the observed in the LIGO / Virgo experiment ratio of the detection rate of black holes to the rate of detection of binary neutron stars requires the assumption of a "conservative" collapse of massive stars into a black hole: almost all the mass of the collapsing star goes under the horizon. This is consistent with the large masses of black holes detected by LIGO/Virgo. On the other hand, the assumption of a small loss of matter during the collapse into a black hole is in good agreement with the small eccentricity of Single-lined Binaries. At the same time, the absence of X-rays from most black holes in binary systems with blue stars is explained. We argue that three sets of LIGO / Virgo observations and data on the Single-lined Binary with a Candidate Black Hole Component confirm the scenario of the evolution of massive field binaries.

**Key words:** gravitational waves: LIGO/Virgo – stars evolution: black holes

## 1 INTRODUCTION

The most populous events detected by LIGO/Virgo gravitational-wave collaboration are merging black holes (Abbott B. et al.2016a, 2016b, 2017, 2019a,b).

In fact, the very discovery of gravitational waves as the black holes merging was predicted in terms of the several scenario of the evolution of massive stars with weak and strong stellar wind dating back to the 1990-ies (Lipunov et al. 1997a, 1997b, 1997c). Tutukov & Yingelson (1993) already pointed out the possibility of such domination.

In the early 1980s, population synthesis of relativistic stars in binary systems was started by the Monte Carlo Scenario Machine (Lipunov, 1982; Kornilov & Lipunov, 1983a,b). This Scenario Machine includes a superposition of two other scenarios developed in parallel in 1970-ies: the evolution of normal stars and the evolution of relativistic stars. The main ideas of the evolution of binaries with normal components were laid down by Paczynski (1971), Tutukov and Yungelson (1973), van den Huevel & Heise (1972), van den Huevel (1994). On the other hand, this scenario includes the evolution of relativistic stars (neutron stars and black holes), which appear in massive binaries during late stages of their evolution (Schwartzman V.F. 1970,1971a,b; Illarionov & Sunyaev 1975; Bisnovatyi-Kogan & Komberg 1975; Shakura, 1975; Wickramasinghe & Whelan 1975; Lipunov & Shakura 1976; Savonije & van den Heuvel 1977; Lipunov 1982; Lipunov 1987; Kornilov & Lipunov 1983a; Lipunov 1992) were the first to develop a combined scenario involving the evolution of relativistic and normal stars and all the relevant literature was discussed in the paper by Lipunov et al. (1996a,b).

The studies involving Scenario Machine (Lipunov et al. 1997a,b,c) provided strong suggestions that black holes will dominate the detected GW transients. Here we report the hypothetical event rates for black-hole and neutron-star mergers in accordance with Lipunov et al. (1997b). The idea of that work was to demonstrate that black holes should be the dominating type of objects detected on a LIGO-type gravitational-wave interferometer irrespectively of the parameters of the scenario of the evolution of binary stars with weak or strong stellar wind. By weak wind, we mean such a loss of mass by a star in the form of a stellar wind that does not change its initial mass by more than 10-30%.[1]

Below, to describe the nature of the conservatism, we use the $k_{BH}$ parameter equal to the fraction of the mass of the pre-supernova star entering the black hole during the collapse. So (1- $k_{BH}$ ) is the mass loss of the presupernova that determines the eccentricity of the resulting binary.

However, the mass-conservative collapse inevitably leads to an increase in the number of black holes in massive binary systems, which could wither as massive X-ray binaries of the Cyg X-1 type. However, we see only a few candidates for such systems in our galaxy. This contradiction is removed in the second paragraph of the letter, where we show why most black holes in massive binaries are not sources of powerful X-rays.

The discovery of black-hole candidates in single-lined FS CMa type systems (Khokhlov, 2018) and LB-1 (Liu et al., 2019 see, however, El-Badry & Quataer,2019; Eldridge, 2019; Simón-Díaz et al., 2020) serves as an excellent confirmation of the classical scenario of the evolution of massive star.

## 2. LIGO/Virgo Black holes: relative detection rate and mass distribution.

After the discovery of merging black holes (Abbot et al. 2016) some authors began to suggest that progenitors of such systems cannot be common field close binary stars. Masses inferred in the LIGO/Virgo experiment proved to be substantially greater than those of the black-holes in most of the observed x-ray binaries. This prompted the researches to suggest that progenitors of merging black holes may be massive Population-III stars deficient in heavy elements and therefore having weak stellar winds (e. g. Mapelli et al. 2009; Zampieri& Roberts 2009; Belczynski et al. 2010; Spera et al. 2015; Belczynski et al. 2016a)[2] . Even more exotic scenarios were proposed to explain the birthnof such pairs in the early cosmological epoch (Dolgov & Postnov 2017).

Here we are analyzing the observed superiority of the detection rate of black holes on the detection rate of binary neutron stars, predicted by Lipunov et al. (1997b).The idea of that work was to demonstrate that black holes should be the dominating type of objects detected on a LIGO-type gravitational-wave interferometer irrespectively of the parameters of the scenario of the evolution of binary stars with weak or hard stellar wind (Fig.1).

The Scenario Machine population synthesis of binary stars with relativistic components (Lipunov et al., 1996b; 1997c) shows that their physical characteristics (separation, eccentricity, component masses) and population size are primarily determined by several key parameters that are still poorly known.

This is due to the lack of an accurate theory of such critical processes as stellar wind, mass exchange between companions, and especially stages with a common envelope. Until now, the asymmetry of the supernova explosion theory remains a mystery. The high level of detail in the evolutionary tracks of single stars, cannot guarantee us the accuracy of determining the characteristics of relativistic binary systems.

---

[1] In other words, we do not go into detail on the relationship between the stellar wind power and the metallicity of stars. It is understood that higher metallicity results in greater mass loss. Of course, the track of a massive star becomes shorter, but this has an insignificant effect on the frequency of mergers, since the initial distribution of binary system separations turns out to be flat (Iben & Tutukov, 1984 ) And wider pairs of progenitors take the place of closer systems.

Taking into account all the theoretical uncertainties in the production of relativistic binaries, we use a simple model of the LIGO / Virgo type detector and show that we still continue to qualitatively correctly explain the observed data. The same goes for the Cosmic star formation history function. We were the first to take into account Star Formation Rate in merging relativistic binary science. Since 1995 (Lipunov et al. 1995).

Since 1995 (Lipunov et al. 1995), we use a simple model in which Universe has first generation stars that formed at z = 2-3 and stars with solar metallicity, which are formed with a constant star formation rate. Yes, this model is certainly inferior to the modern Madau function (Madau & Dickinson, 2014), which we used in later works in cases where its knowledge was decisive. For example, describing the evolution of Supernova Type Ia Rate for elliptical galaxies in the model of merging white dwarfs (Lipunov, Panchenko & Pruzhinskaya, 2011). For the merging of relativistic stars (the birth history of which contains two supernova explosions), the exact Madau function is little consolation compared to the dark parameters of the evolution of binary stars, and we used a more crude model of star formation in the Universe. As a result, we not only received high-quality predictions within the framework of a simple classical scenario, but we also continue to explain and predict new observable properties of Single-lined Binaries with a Black Hole Candidate Component.

Therefore, these uncertain understanding have been parameterized in the calculations by introducing the following parameters: wind power as a fraction of the mass of a star lost by it at each stage (weak and strong stellar wind), parameter common shell which shows the efficiency of the selection of the orbital momentum at the stage of the common shell and kick velocity obtained during the formation (collapse) of a relativistic star. For black holes, it is necessary to introduce an additional parameter $k_{BH} = M_{BH} / M_{pre}$ - the ratio of the black hole mass to the pre-supernova mass. Obviously $0 < k_{BH} \leq 1$. In addition, it is necessary to maintain a critical initial (zero age) mass of a star capable of generating black holes $M_{min}$ and introduce kick velocity, which a black hole acquires in the process of star formation. We relied on data from observations of neutron stars - radio pulsars for which the average kick velocity $W_{NS}$ = 200 km / s is distributed according to the Maxwell function (Lipunov, 1997c). Of course, one should expect that the kick velocity for black holes $W_{BH}$ will be lower on average, since black holes are "heavier". We have suggested that the anisotropy is proportional to the mass of the ejected shell, that the following phenomenological formula for the kick velocity of a black hole is (Lipunov, 1997c):

$$W_{BH} = W_{NS} (1 - k_{BH}) / (1 - M_{OV} / M_{pre}),$$

where $M_{OV}$ is the Oppenheimer-Volkov limit. If $M_{BH} \to M_{OV}$ then $W_{BH} \to W_{NS}$

Conversely, if $k_{BH} \to 1$ then $W_{BH} \to 0$. The aim of Lipunov et al. (1997b) showed that, despite the uncertain understanding in the evolution of binary systems, black holes will prevail on LIGO / VIRGO interferometers. Therefore, in the beginning, we simply "let go" of all evolution parameters in scenarios of strong and weak stellar winds. We have modernized the graph from this old work by plotting relative values along the vertical axis and added the results of observations in the LIGO / Virgo experiment (Fig. 1).

The large blue domain in Fig.1 corresponds to the normalized black-hole event detection rate as a function of the $k_{BH}$ for arbitrary parameters of SN anisotropy, common-envelope efficiency, initial distribution of component mass ratio, etc. We emphasize that the parameters were scattered within mathematically permissible limits without paying attention to the observations. . The green area, that is reminiscent of the Loch Ness Monster's head corresponds two restrictions. The first limitation is due to the absence of binary radio pulsars with black holes (Lipunov et al. 1995). This imposes an upper constraint on the possible black-hole merger rate. On the other hand, there should be several blue massive stars with black holes (of the Cyg X-1 type) in our artificial Galaxy and this condition imposes a lower constraint on the merger rate. Recall that the detection rate of mergers involving black holes officially confirmed by the LIGO/Virgo team is about one order of magnitude higher than the detection rate of events involving neutron-star mergers and is statistically estimated at R(BH)/R(BNS) ~ 1:10 (Abbott et al. 2019). This ratio is approximately confirmed by the ongoing O3 observations performed on LIGO/Virgo. According to recently published data, among 44 events, only one event is associated with binary neutron star mergers (Abbott et al. 2020b). Thus, the observational value in all confirmed data we take R (BH) / R (BNS) ~ 49: 2. Then a comparison of the predicted merger rate with observational one requires a very high value of $k_{BH} \approx 0.9$-$1.0$ (Fig.1).

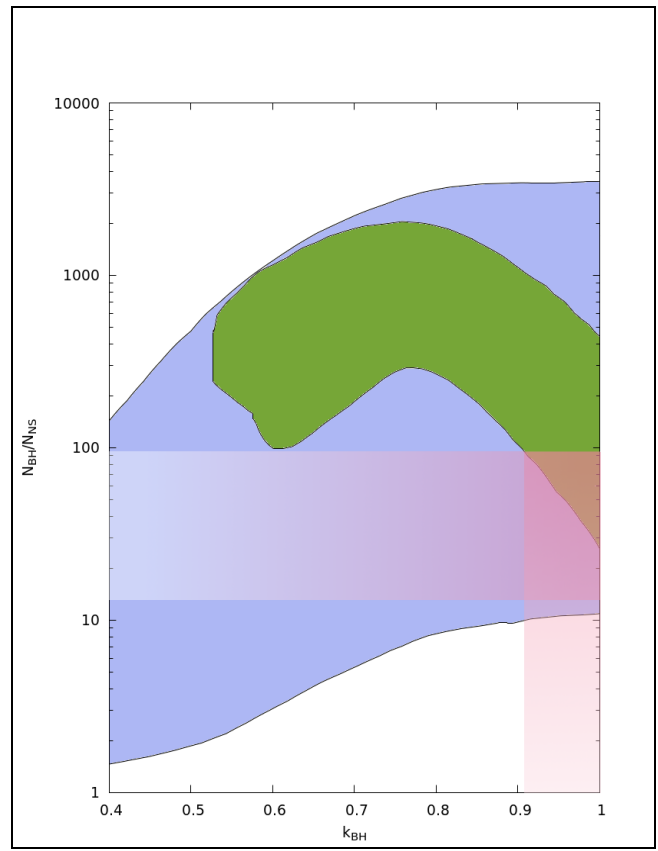

**Fig 1**. Loch Ness Monster diagram. Relative detection rate of gravitational-wave events on LIGO/Virgo type interferometers as a function of the fraction of the mass of the presupernova remaining in the black hole in the case of the scenario with weak and strong stellar wind (Lipunov et al. 1997b). The large blue domain corresponds to the normalized black-hole event detection rate as a function of the $k_{BH}$ for full arbitrary parameters of SN anisotropy, common-envelope efficiency, initial distribution of component mass ratio, etc. The "head of the monster" (green area) corresponds to the domain of detection rates of events involving black noles for the most likely parameters of the evolutionary scenario (see text and Lipunov et al., 1996a,b). The horizontal strip corresponds to the LIGO/Virgo statistics computed based on the observed event ratio BH/BNS = 49:2 ratio (Abbott et al. 2020b).

Obviously, mass conservation collapse increases the expected masses of black holes, which is in qualitative agreement with the LIGO / Virgo data. To illustrate this effect, we carried out a population synthesis by Scenario Machine of merging black holes within the framework of one specific evolutionary scenario with a weak stellar wind. The specific parameters of this scenario are listed in Tables 1 and 2 and discussed in detail in the next section.

Almost immediately, we obtained good qualitative agreement with the observed masses of merging black holes. In particular, our simulations contained a significant number of mergers with a total mass of more than 100 solar masses. Note that such massive stars are born as a result of pair instability (Woosley, Blinnikov & Heger, 2007) and for them everything ends in a supernova explosion and such a collapse can hardly be called quiet. Nevertheless, Woosley (2017) notes that the boundary for unstable collapse is not very well defined and can reach 80 solar masses! In this case, it is quite possible to expect a merger of black holes with a total mass of up to ~100 solar masses (see Abbott et al. 2020a).

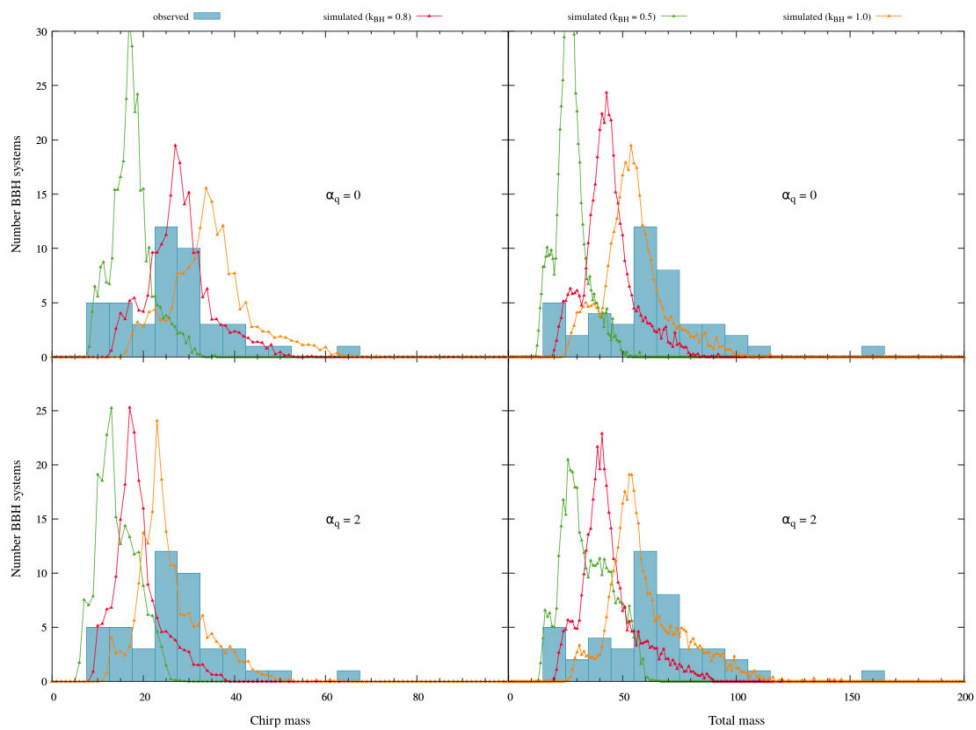

**Fig 2**. **The distribution function of merging black holes by chirp masses (left panel) and by total mass (right panel).** Blue are observational data (Abbott et al. (2019), Abbott et al. (2020b)). The results of population synthesis with low wind scenario are corrected for observational selection, three diagrams in each panel for three values of the fraction of mass that collapses into a black hole $k_{BH}$ = 0.5, 0.8 and 1.0. Top panel - scenario with a flat distribution of mass ratio function $dN/dq = q^\alpha$. $\alpha_q = 0$. Lower - $\alpha_q = 2$.

The most accurately determined quantity in the LIGO / Virgo experiment is the so-called chirp mass. The results of gravitational wave observations of the first three series O1, O2 and O3 were recently published in Abbott et al. (2019,2020b). It is these results that we will try to compare with population synthesis within the Scenario Machine. It should be emphasized that gravitational-wave experiments primarily register mergers of black holes with large chirp mass. This leads to the effect of observational selection, which we took into account in the population synthesis data using the method from Lipunov et al. (2017).

Here we briefly recall that the main selection effect is associated with the fact that the sensitivity of gravitational-wave interferometers of the LIGO / Virgo type is proportional to the chirp mass $\sim M^{5/6}$. Accordingly, the volume inside the horizon of the detector's sensitivity increases as $\sim M^{15/6}$. Thus, the simulated rate distribution of merging black holes in galaxies is reduced to a detectable event rate.

In Figure 2. we present the results of observations of O1, O2 and O3 population synthesis for three $k_{BH}$ values. Obviously, the best agreement is achieved at $k_{BH} => 1$. This is also illustrated by the calculated functions of the total mass of merging black holes (right panel in Fig. 2). Read more about the scenario parameters in the next section.

Analyzing the observational and model results in Fig. 2, we must understand that we are dealing with small statistics of experimental data and uncertainty by hidden parameters of the theoretical scenario. Therefore, we did not try to find an exact match. However, it is obvious that the best agreement between observations and the results of population synthesis is achieved for scenarios in which the formation of black holes in close binaries proceeds with a large conversion of mass into a black hole.

### 3. Simulation of blue stars in a pair of black holes.

As we have already mentioned in recent years, a whole class of blue stars with single spectral lines has been discovered (Khokhlov 2018), which may indicate the presence of black holes as the second component. In the case of the LB-1 system, the presence of an anomalously large mass in the black hole system was even reported (Liu et al. 2019). However, this result has been subject to much criticism (El-Badry & Quataert, 2019; Eldridge, 2019; Simón-Díaz et al., 2019). It is interesting that in all cases we were talking about very small orbital eccentricities of these orbits. Therefore, we undertook new calculations of population synthesis, which unexpectedly gave good agreement of gravitational wave data with observations of optical astronomy. Indeed, as we have shown in the previous section, the best agreement of the classical population synthesis of the Scenario Machine is achieved for the so-called silent collapse, in which practically the entire mass of the pre-supernova goes into the black hole (Fig. 1.). How can this manifest itself at earlier evolutionary stages, for example, at the stage after the first mass exchange and the formation of a black hole. Indeed, it is these systems that are now found in spectral observations. The growth of $k_{BH} \to 1$ leads to two striking consequences. The first is that the absence of the dumped mass during the collapse leaves the orbit practically circular. And second, all other things being equal, an increase in $k_{BH}$ will inevitably lead to an increase in the masses of merging black holes, as in LIGO / VIRGO observations (see section 2).

To illustrate the first consequence of "quit" collapse, we performed a new population synthesis of specially blue stars with black holes.

We used the Scenario Machine to perform population synthesis for the optimum parameter values that we inferred in our earlier study (Lipunov et al. 1996a,b; Bogomazov & Lipunov 2008).

For our computations we used the online version of the Scenario Machine to simulate and analyze $\sim 10^7$ systems with different $k_{BH}$.

In these calculations, we chose one of the scenarios for the evolution of stars with a weak stellar wind, which naturally explains the large total mass of blacks observed in the LIGO/ Virgo experiments

A very important parameter of population synthesis is the initial mass ratio function. We usually used several initial distribution functions. In calculating the head of the Loch Ness Monster (Lipunov et al. 1997b), various distribution functions are used. Naturally in 1997 we could not use modern distribution functions such as Moe & Di Steffano (2014). Such detailed data from a large study of binary stars undoubtedly move forward our understanding of the evolution of binary stars at a stage before the formation of relativistic companions. However, when modeling the eccentricities of relativistic stages with, due to uncertainties in the knowledge of the hidden evolution parameters, this overestimated accuracy turns out to be ineffective. Moreover, using simpler parameterized initial distributions and me in the process of population synthesis, we can understand how these initial functions change the final result of the evolution of a massive binary system. It should be emphasized that we have extended our calculations for two values of the initial mass ratio function. It should be said that in the calculations of 1997 we have already used the initial functions with a maximum at $q = 1$ ($f(q) \sim q^2$.) We have included this function in the calculations of eccentrists and masses. In the end, we tried to show that single-lined binaries with a Black Hole candidate component should be round.

Table 1 lists the parameters and their distribution functions used to generate simulated systems. These are the main parameters affecting the evolution of the binary: the component masses $M_1 > M_2$, semimajor axis of the system (a), and the a priori unknown coefficient $k_{BH}$ used to construct the distributions. The other parameters (see Table 2) were fixed and remained unchanged in the simulation process. More detailed parameters can be found on the open site of the Script Engine (Nazin et al.,1998).

**Table 1.** Boundaries and distribution of the parameters varied in the process of simulation

| Parameter | Distribution | Lower boundary | Upper boundary |
|---|---|---|---|
| Initial Mass ($M_1$) | $\sim M^{-2.35}$ | 10 | 120 |
| Initial mass ratio in the binary ($q = M_2/M_1$) | $\alpha_q = 0, 2$  $dN/dq \sim q^{\alpha_q}$ | 0 | 1 |
| Semimajor axis (a) | $dN/da \sim a^{-1}$ | 10 $R_\odot$ | $10^8 R_\odot$ |
| collapse mass fraction | $k_{BH}$ | 0 | 1 |

**Table 2**. Parameters of the Scenario Machine[3] that remain unchanged during the simulation process.

| e - initial orbital eccentricity | 0 |
|---|---|
| T - evolution time (yr) | 1.5e10 |
| Kick velocity (km/s) $W_{NS}$ | Maxwellian, 200 |

| Kick direction | Uniform random direction |
|---|---|
| Normal star mass loss: | Low 10% |
| Maximal accretion rate into CE: | Eddington |
| Matter acception by normal star during accretion: | Partically nonconsercative |
| Common envelope efficiency | 1 |
| $M_{min}$ - Initial mass for BH formation | 35 $M_\odot$ |
| Oppenheimer-Volkoff limit | 2.5 $M_\odot$ |

We then selected among the resulting tracks those where one of the intermediate evolutionary stages coincided with the chosen one. In particular, to construct the distribution shown in Fig. 2, we chose a bound binary system consisting of a black hole and a main-sequence star. We analyzed the data for such systems separately(see section 2).

In Fig. 3 we show the eccentricity of the binary system at the black-hole – main-sequence star stage plotted versus $k_{BH}$. The black line indicates the average eccentricity of binary systems with the given $k_{BH}$, whereas the violet region shows the domain occupied by 90% of systems with the given $k_{BH}$. An example of a track passing through the given stage can be found at Nazin et al. (1998). The wide violet region shows the limits of the orbital eccentricity variation after the formation of the black hole, depending on the fraction of the dropped mass and the parameters of the binary system's orbit.

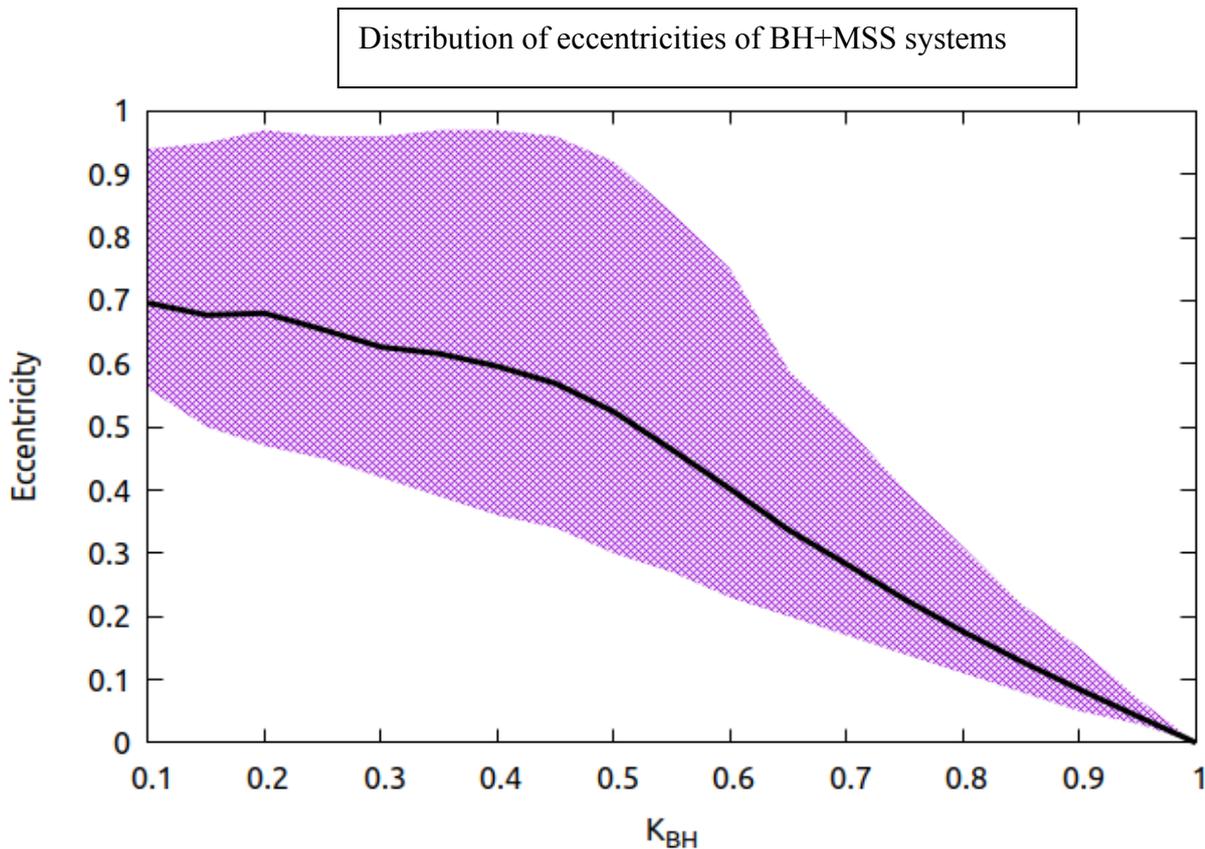

**Fig. 3. Distribution of an eccentricity as a function of $k_{BH}$ for Black Hole +Main Sequence Star systems.**
The black line indicates the average eccentricity of binary systems with the given $k_{BH}$, whereas the violet region shows the domain occupied by 90% of systems with the given $k_{BH}$.

So, within the framework of the scenario under consideration, the observed LIGO / Virgo relative rates of merging of black holes and neutron stars require almost complete conservation of mass during a supernova explosion ($k_{BH} \to 1$). However, this means that the orbits of black holes after the explosion of the first supernova should be almost circular, which we see in Fig. 2. It is these systems that have been discovered in recent years (Khokhlov et al. 2018).
Finally, we will discuss the third consequence of our scenario in the next part.

## 4. Why black holes paired with blue stars in detached binaries are not X-ray binaries?

The scenario we are discussing with a low mass loss during the formation of black holes ($K_{BH} \to 1$) leads to more frequent formation of black holes in a system with blue stars. As a consequence, we should have expected a large number of X-ray sources in the Galaxy. Of special interest is the question why single-lined binaries with a candidate black hole component (Khokhlov et al. 2018) are not X-ray binaries? Note, that Chandra X-ray Observatory detected no x-ray flux from LB-1 and imposed an upper limit on the x-ray luminosity of the system $< 2 \; 10^{31}$ erg s$^{-1}$(Liu et al. 2019).

Karpov & Lipunov (2001) specially investigated this question posed in the title of their paper: "Why Do We See So Few Black Holes in Massive Binaries"? The problem was that the most bona fide black-hole candidate at the time - Cyg X-1 - was a very massive binary where the blue star practically fills its Roche lobe. This is a very short-lived stage of the evolution of the binary system. However, it is in this case that the matter captured by the black hole forms an accretion disk converting gravitational energy into x-ray emission with very high efficiency (Shakura 1972). It is, however, clear that there should be tens and hundreds times more wider systems. But no such systems are found among x-ray binaries! Where are they?

The answer was obtained: first, accretion disk does not form in such systems and quasi-spherical type of advection-dominated accretion is realized. The condition of the formation of accretion disk around the black hole in the case of accretion from the stellar wind can indeed be fulfilled only if the specific angular momentum $k$ exceeds the angular momentum at the last stable orbit ($r_{min} = 3\, r_g/c^2$):

$$k > \sqrt{GM_{bh}\, r_{min}} \qquad (1).$$

According to Illarionov & Sunyaev (1974), $k \approx (1/4)\, \Omega\, r_B^2$. We thus obtain the condition of the formation of accretion disk for circular orbits (cf Lipunov 1992):

$$V_w < Vcr \approx 250\ km/s\ m_{10}^{1/4}\ T_{100}^{-1/4}\ (1+tg^2\beta)^{-1/2} \qquad (2)$$

where $m_{10} = M_{bh}/10 M_\odot$, $T = T/100^d$, $tg\beta = V_{orb}/V_w$.

Given that the typical velocity of radiative wind at a distance of 2-3 stellar radii is $V_w \sim 1000 - 2500$ km/s, it is evident that the disk can form only in very close systems of the Cyg X-1 type where the stellar wind has not yet accelerated to high velocities.

Thus quasispherical accretion with extremely strong advection is the process that occurs in the bulk of the matter of binary systems consisting of a black hole and a blue star (Narayan & Yi., 1995). Magnetic field in the accreted flow could increase the efficiency of the energy release (Shvartsman, 1970b), but Karpov & Lipunov (2001) showed that the "magnetic pumpout" mechanism (Lipunov, 1992) operates twice in such systems and as a result, magnetic field plays no part in the accretion flow.

Let us now compute the effect of magnetic pumpout. We assume that near the surface of the blue star magnetic energy is equal to the thermal energy of the stellar-wind plasma $\varepsilon_m(0) = \varepsilon_{th}(R0)$. Actually, $\varepsilon_m(0) < \varepsilon_{th}(R0)$ in all normal stars except the so-called magnetized ones. The thermal energy is equal to $\varepsilon_{th} \approx \rho V_s^2$, where $V_s$ is the sound speed in the matter. We further assume that magnetic field lines are totally frozen in the stream of spherically symmetric wind $4\pi R^2 B = const$, to obtain, at the distance of the semimajor axis $a$, $\varepsilon_m(a) = B^2/8\pi = \varepsilon_m(0)(Ro/a)^{-4}$.

Let us now compare the magnetic-field energy with the kinetic energy density in the stellar wind: $\varepsilon_{kin}(a) = \rho V^2/2 = \dot{M}\, Vwin\, /8\pi R^2 \sim R^{-2}$.

Thus the magnetic-field to kinetic energy ratio at the distance of the semimajor axis is:

$$\varepsilon_m(a)/\varepsilon_{kin}(a) = \mathcal{M}^{-2}\, (R_0/a)^{-2}. \qquad (3)$$

where $\mathcal{M} = V_{win}/V_{sound}$ is the Mach number. Here we assume that the temperature of the stellar wind remains unchanged and is close to the surface temperature of the star. This is approximately true because stellar winds of hot blue stars are Stroemgren zones (Mihalas 1978). However, at the gravitational-capture radius (the Bondi radius) $r_B = 2GM/V_{win}^2$ specific kinetic energy of the wind is equal to specific gravitational energy. Thus if at the base of the stellar wind magnetic energy is equal to thermal energy then at the orbit of the black hole the former may, in turn, be much smaller than the gravitational energy! This sort of magnetic pumpout operates twice – first, due to the high velocity of the stellar wind and second, due to diverging stellar-wind flow lines, which make the magnetic-field structure radial. After a small part of the stellar wind is captured the field lines again become radial, but now they are directed toward the black hole and the role of the magnetic field increases:

$$\varepsilon_m / \varepsilon_g = \mathcal{M}^{-2} (R_0/a)^{-4} (r/r_B)^{-4}. \qquad (4)$$

The typical velocity of radiative stellar wind in the case of LB-1 type stars is $V_{win} \sim 1000\text{-}2500 \text{km/s}$. In particular, the Bondi radius for the stellar-wind velocity of 2500 km/s is equal to $r_B \approx (10\text{-}15) R_\odot \ll a$. As a result, magnetic-field energy remains smaller than gravitational energy throughout the entire accretion flow down to the event horizon. This means that the energy release efficiency of such a black hole is negligible. That is why wide pairs made of a black hole and a blue star do not become x-ray binaries (Karpov & Lipunov 2001).

## 5. Discussion.

Thus we showed that within the classical scenario of the evolution of binaries with an weak stellar wind naturally explains not only the observed parameters of single-lined binaries with a candidate black hole component and results of LIGO/Virgo experiment (Abbott et al. 2019). Moreover, the main facts of both experiments - namely, domination of black holes, and their particular fraction were actually predicted in early Scenario Machine computations.

The usual objections against the results obtained with the Scenario Machine are that these computations purportedly operate with very approximate nuclear evolution scenarios developed in the 20th century. Modern population-synthesis computations are believed to take into account the progress of our understanding of the evolution of single and binary stars (Moe M. & Di Steffano R., 2015) and this fact alone does not allow us to take into account the computations and models from 20 years ago (see for example Belczynski et al., 2016b). However, it should be noted that until recently new codes did not include the evolution of relativistic stars in spin calculations. Recall that these were the first Scenario Machine codes that included not only nuclear evolution, but also spin evolution of magnetized neutron stars (Kornilov & Lipunov 1983a, b). Of course, taking into account the spin evolution of magnetized neutron stars does not directly affect the calculations of the merging rate of black holes. But, if we recall the hidden undefined parameters - for example, kick velocity, or the efficiency of the common envelope stage, or the stellar wind power, then everything changes. After all, we determine the hidden parameters after comparison with observations. Latent parameters can be determined from systems with neutron stars, for example, by comparing the number of single and binary radio pulsars (see Lipunov et al. 1997c). In addition, neutron stars are observed at completely different stages of the evolution of the binary system (Lipunov et al., 1996), and here nothing can be determined without knowing the spin evolution.

Fortunately, recently, works began to appear in which the spin effects of the evolution of magnetized neutron stars and white dwarfs have finally been included in the evolution of binary stars (for example Chattopadhyay et al. 2020).

And, secondly, however accurately the nuclear tracks of single stars or spin evolution are computed (which is also highly questionable), there is hardly any progress in our understanding of the evolution of binaries because of the uncertainty of the stages in the development of binary systems that are crucial for the fate of their components: the common-envelope stage, kick velocity during supernova explosion.

The authors of alternative scenarios seek evolutionary tracks that could somehow explain the discovery of non-x-ray binaries a with black hole and such a scenario indeed became highly popular in some population-synthesis models (Liu et al. 2019). Here we deal with the possible binary nature of the "enormous" mass of the black hole in this system by a merger of two black holes ( Abbott et al. 2020; LIGO Scientific Collaboration and Virgo Collaboration, 2020; Sakstein et.al, 2020).

The some researchers continue to consider exotic scenarios for the formation of black holes and neutron stars detected in gravitation-wave experiments (Abbott et al. 2019a) – this is collision capture of stars in globular clusters - or even invoke the cosmological nature of massive black holes. Let us note here a number of scenarios that have been discussed recently. The birth of binary black holes in star clusters (Antonini et al., 2020), in triple systems (Fragione and Loeb, 2019), birth from quadruples (Fragione and Kocsis, 2019), the birth of pairs of large black holes in the process of collapse caused by pair- instability (Spera et al., 2019; Carlo et al., 2020, also see Woosley et al., 2007).

Here we note that the event GW190521 is at the farthest edge of the total mass distribution of mergers (Fig. 2b). Perhaps this requires the involvement of the trinity of the parental system (Fragione and Loeb, 2019).

Of course, we do not deny progress in understanding and calculating the evolutionary tracks of single stars. However, we emphasize once again that in the life of binary systems, such parameters as nuclear collapse and supernova explosion (kick velocity), the common envelope stage, non-conservative exchange of mass and rotational moment remain much more uncertain and more critical. Added to this are the hitherto unclear hydrodynamics of the radiative stellar wind (clumping problem).

This, in fact, negates our progress in understanding some of the initial stages of the evolution of binary systems with normal stars (refinement of tracks at the stage of single stars, detailed consideration of metallicity, etc.). It is better to parameterize and determine them by comparing the characteristics of relativistic stars with observations, much more important parameters associated with short and stormy stages in their life.

## 6. Conclusions

In this note, we have shown that, within the framework of the classical scenario of stellar evolution, the data from the LIGO / VIRGO experiments showing the dominance of black holes and the new discoveries of single-lined sinaries with a black hole candidate component agree on the assumption that practically all the mass of the progenitor core goes into black during the collapse. hole ($k_{BH} \rightarrow 1$). In this case, the orbits of such binaries remain circular even in fairly wide separated pairs. The second consequence is the birth of anomalously massive pairs of stellar mass black holes.

"Conservative" or "quiet" collapse is characteristic of those binary systems in which progenitors have almost completely lost their hydrogen envelope during the first mass exchange. Single massive stars with a weak stellar wind in the process of collapse can shed their hydrogen envelopes, leading to the luminal SN phenomenon.

This is a key deficit of our paper there is a lot of very broad comparisons to broad observed facts but no hard detailed comparison between numbers that are well known. But in this short note we tried to draw attention to the fact that we can try to clarify the dark parameters of the physics of the formation of black formations of black holes by comparing so many different data of optical spectroscopy and only incipient gravitational-wave astronomy.

We believe that after the LIGO/Virgo discovery (Abbott et al. 2016a, 2017, 2019) of the first merging black holes predicted by Scenario Machine in 1997 (Lipunov 1997a,b), merging neutron stars (Lipunov 1987; Lipunov & Pruzhinskaya 2013), and the discovery of mixed pairs in the observing sets O1, O2, O3[**] (more then 50 events! in O1-O3 sets), it becomes clear that the progenitors of all these objects belong to the same population of the Universe - massive field binaries with weak stellar wind.

However, we do not think that this can be done given only one code, the consistency of the binary evolution should be achieved by considering the output of a series of codes.

ACKNOWLEDGEMENTS

We are grateful to the anonymous referee for the tough objective review.
VL is supported by RFBR BRICS grant 17-52-80133 grant. EG is supported by RFBR 19-29-11011. In conclusion, we thanks S.I.Blinnikov, A.M.Cherepashchuk, A.V.Tutukov, A.I.Bogomazov, K. Zhirkov, A.Chasovnikov and all the participants of the Zel'dovich Joint Astrophysical Seminar OSA.